\documentclass[
reprint,
superscriptaddress,
amsmath,amssymb,
aps,
pra,
floatfix,
]{revtex4-2}

\def\final{0}

\usepackage{braket}
\usepackage{mathtools}
\usepackage{graphicx}
\usepackage{color, xcolor}
\usepackage{dcolumn}
\newcolumntype{z}[1]{D{.}{.}{#1}}
\usepackage{physics}
\usepackage{upgreek}
\usepackage{subfigure}
\usepackage{tabularx}
\newcolumntype{Y}{>{\centering\arraybackslash}X}

\usepackage[linesnumbered,ruled ]{algorithm2e}
\SetKwInput{KwOutput}{Output}
\SetKwInOut{Input}{Input}
\SetKwInput{Output}{Output}
\UseRawInputEncoding 

\usepackage{xargs}
\usepackage[colorinlistoftodos,prependcaption]{todonotes}

\definecolor{Asparagus}{rgb}{0.53, 0.66, 0.42}
\definecolor{cornflowerblue}{rgb}{0.39, 0.58, 0.93}
\definecolor{darkolivegreen}{rgb}{0.33, 0.42, 0.18}
\definecolor{awesome}{rgb}{1.0, 0.13, 0.32}

\newtheorem{definitionenv}{Definition}
\newtheorem{lemmaenv}[definitionenv]{Lemma}
\newtheorem{theoremenv}[definitionenv]{Theorem}
\newtheorem{corollaryenv}[definitionenv]{Corollary}
\newtheorem{propositionenv}[definitionenv]{Proposition}
\newtheorem{conjectureenv}[definitionenv]{Conjecture}
\newtheorem{remarkenv}[definitionenv]{Remark}
\newenvironment{remark}{\begin{remarkenv}\rm}{\end{remarkenv}}
\newcommand{\br}{\begin{remark}}
	\newcommand{\er}{\end{remark}}

\newtheorem{exampleenv}{Example}
\newtheorem{app-lemmaenv}[section]{Lemma}

\newenvironment{definition}{\begin{definitionenv}\rm}{\end{definitionenv}}
\newenvironment{lemma}{\begin{lemmaenv}\rm}{\end{lemmaenv}}
\newenvironment{theorem}{\begin{theoremenv}\rm}{\end{theoremenv}}
\newenvironment{corollary}{\begin{corollaryenv}\rm}{\end{corollaryenv}}
\newenvironment{example}{\begin{exampleenv}\rm}{\end{exampleenv}}
\newenvironment{proposition}{\begin{propositionenv}\rm}{\end{propositionenv}}
\newenvironment{conjecture}{\begin{conjectureenv}\rm}{\end{conjectureenv}}
\newenvironment{app-lemma}{\begin{app-lemmaenv}\rm}{\end{app-lemmaenv}}

\newcommand{\bd}{\begin{definition}}
	\newcommand{\ed}{\end{definition}}
\newcommand{\bl}{\begin{lemma}}
	\newcommand{\el}{\end{lemma}}
\newcommand{\elp}{\hspace*{\fill} $\Box$
\end{lemma}}
\newcommand{\bt}{\begin{theorem}}
\newcommand{\et}{\end{theorem}}
\newcommand{\etp}{\hspace*{\fill} $\Box$
\end{theorem}}
\newcommand{\bc}{\begin{corollary}}
\newcommand{\ec}{\end{corollary}}
\newcommand{\ecp}{\hspace*{\fill} $\Box$
\end{corollary}}
\newcommand{\bcj}{\begin{conjecture}}
\newcommand{\ecj}{\end{conjecture}}

\newcommand{\be}{\begin{example}}
\newcommand{\ee}{\end{example}}
\newcommand{\eep}{\hspace*{\fill} $\Box$
\end{example}}
\newcommand{\bp}{\begin{proposition}}
\newcommand{\ep}{\end{proposition}}
\newcommand{\epp}{
\end{proposition}}

\ifnum\final=0
\newcommand{\mynote}[2]{{\color{#1} \marginpar{\tiny #2}}}
\newcommand{\mybignote}[2]{{\color{#1} $\langle \langle$ #2$\rangle \rangle$}}
\newcommandx{\rednote}[2][1=]{\todo[linecolor=red,backgroundcolor=red!25,bordercolor=red,#1]{#2}}
\newcommandx{\bluenote}[2][1=]{\todo[linecolor=blue,backgroundcolor=blue!25,bordercolor=blue,#1]{#2}}
\newcommandx{\yellownote}[2][1=]{\todo[linecolor=yellow,backgroundcolor=yellow!25,bordercolor=yellow,#1]{#2}}
\newcommandx{\greennote}[2][1=]{\todo[inline,linecolor=olive,backgroundcolor=green!25,bordercolor=olive,#1]{#2}}

\else
\newcommand{\mynote}[2]{}
\newcommand{\mybignote}[2]{}
\newcommand{\rednote}[2][1=]{}
\newcommand{\bluenote}[2][1=]{}
\newcommand{\greennote}[2][1=]{}
\newcommand{\yellownote}[2][1=]{}

\fi

\usepackage{adjustbox}
\usetikzlibrary{quantikz2}
\usetikzlibrary{backgrounds,fit,decorations.pathreplacing}  
\usetikzlibrary{automata} 
\usetikzlibrary{circuits.logic.CDH}
\usetikzlibrary{circuits.logic.US}
\usetikzlibrary{mindmap}
\usetikzlibrary{decorations}
\usetikzlibrary{decorations.pathmorphing}
\usetikzlibrary{arrows,decorations.pathreplacing}

\tikzset{meter/.append style={draw, inner sep=10, rectangle, font=\vphantom{A}, minimum width=30, line width=.4, path picture={\draw[black] ([shift={(.1,.3)}]path picture bounding box.south west) to[bend left=50] ([shift={(-.1,.3)}]path picture bounding box.south east);\draw[black,-latex] ([shift={(0,.1)}]path picture bounding box.south) -- ([shift={(.3,-.1)}]path picture bounding box.north);}}}

\begin{document}

\title{
Overlapped-repetition Shor codes achieving fourfold asymptotic rate
}

\author{En-Jui Chang}
\email{phyenjui@gmail.com}
\affiliation{Taichung 421786, Taiwan}

\date{\today}

\begin{abstract}
Introducing controlled overlap among a few repetition blocks yields a fourfold asymptotic rate improvement while preserving an average stabilizer weight of \(4\). Substituting the overlapped outer layer with an LDPC code further produces a family of constructions with asymptotic rate \(2/d\). We also describe a constant-excitation variant that suppresses collective coherent errors without additional overhead, as well as a bosonic generalization that extends the framework to oscillator encodings. The resulting family achieves a code rate intermediate between that of the rotated surface code (average stabilizer weight \(4\)) and the BB code (average stabilizer weight \(6\)), while remaining free of the performance degradation typical of iterative decoders.
\end{abstract}

\maketitle

\section{Introduction}

The implementation overhead of a quantum error-correcting code (QECC) is a central practical constraint, independent of the target code distance. For a device requiring a fixed distance and a prescribed number of logical qubits to support useful quantum applications, a higher code rate directly reduces the physical-qubit resources. This reduction is particularly valuable for experimental platforms operating under stringent hardware and budgetary limitations. Notably, the bivariate bicycle (BB) codes developed by the IBM Quantum team~\cite{Bravyi2024} achieve an error threshold of \(0.7\%\) while offering a code rate roughly an order of magnitude higher than that of the surface code~\cite{bravyi1998quantumcodeslatticeboundary, Dennis_2002}, which underlies recent demonstrations by Google Quantum AI~\cite{Google2023, GoogleSurface2024}.

Beyond code rate, the stabilizer weight is an equally important design parameter. As emphasized by Gidney and Bacon~\cite{gidney2023baconthreshold}, stabilizers of larger weight are more vulnerable to measurement faults. Even for the canonical Shor code~\cite{Shor95}, Nguyen \emph{et al.}~\cite{Nguyen_2021} experimentally showed that extracting two weight-\(6\) stabilizers associated with degenerate \(Z\) errors introduces more noise than measuring six weight-\(2\) stabilizers for \(X\) errors. This sensitivity partly explains the strong performance of the surface code~\cite{bravyi1998quantumcodeslatticeboundary, Dennis_2002}, whose stabilizers have weight at most four, as demonstrated in recent Google experiments~\cite{Google2023, GoogleSurface2024}. While BB codes substantially improve the code rate, they do so at the expense of increased average stabilizer weight, from about \(4\) in the surface code to \(6\) in BB codes, and current experimental results~\cite{wang2025demonstrationlowoverheadquantumerror} indicate that BB codes with larger distance have not yet realized clear logical-error suppression.

Moreover, increased stabilizer weight leads to higher decoding complexity for belief-propagation (BP) based decoders, whose cost scales linearly with the average check weight when processing syndrome data~\cite{Kuo2022} and leads more measurement errors~\cite{PRXQuantum.5.040302}. This affects both the BP + ordered statistics decoding (OSD)~\cite{PhysRevResearch.2.043423} employed by the IBM Quantum team~\cite{Bravyi2024} and the sliding-window decoder used at NVIDIA~\cite{gong2024lowlatencyiterativedecodingqldpc}, as each window invokes BP. The resulting complexity is reflected in NVIDIA's report that their implementation ``\dots achieves a worst-case decoding latency of \(3\text{ ms}\) per window for the \([[144,12,12]]\) code.'' For comparison, Google notes that ``the qubits have a mean operating \(T_1\) of \(68\,\mu\text{s}\)" and that they perform ``real-time decoding, achieving an average latency of \(63\,\mu\text{s}\)." Even allowing for hardware differences, the roughly \(47\times\) larger latency highlights a substantial performance gap and may constitute a significant bottleneck for reaching subthreshold logical error suppression.

Beyond the aforementioned concerns regarding decoding complexity, iterative decoders can also suffer performance loss when certain error syndromes cannot be resolved within practical time limits. In particular, trapping sets~\cite{Raveendran2021trappingsetsof} may prevent the decoder from identifying a valid recovery operator. For a generic (possibly random) \([[n,k,d]]\) code, the syndrome space contains \(2^{\,n-k}\) elements, and the absence of exploitable structure makes decoding increasingly difficult as this space grows exponentially. Consequently, proposed general-purpose decoders may struggle to reliably handle all syndromes in realistic scenarios.

A straightforward strategy for mitigating such performance loss is to revisit codes with highly regular structure. For example, a \([[d^{2},1,d]]\) Shor code~\cite{PhysRevA.52.R2493, Shor95} contains \(d(d-1)\) weight-\(2\) \(ZZ\) stabilizers and \((d-1)\) weight-\(2d\) \(X^{\otimes 2d}\) stabilizers, requiring only \((1+d)2^{\,d-1}\) recovery rules rather than the full \(2^{\,d^{2}-1}\) syndrome space. Structurally, the Shor code is built from an outer \([d,1,d]\) repetition code to protect against phase-flip errors and an inner \([d,1,d]\) repetition code to protect against bit-flip errors. The average stabilizer weight is \(\tfrac{4d}{d+1} \le 4\), ensuring modest measurement overhead. The drawback, however, is the substantial qubit overhead required to achieve larger distances, resulting in a poor code rate. This trade-off motivates the search for more efficient designs that improve rate while preserving these advantageous structural features.

Indeed, the outer code can be generalized to yield a high-rate amplitude-damping \(w\) code with parameters \([[(w+1)(w+k),\,k]]\)~\cite{PhysRevA.111.052602}, capable of correcting all amplitude-damping errors of weight up to \(w\). Such a construction is intrinsically error-biased. This observation suggests that the Shor code can be extended to encode multiple logical qubits simultaneously. The generalization of the outer code in Ref.~\cite{PhysRevA.111.052602} was itself inspired by the scheme of Fletcher \emph{et al.}~\cite{4675715}, where the outer code is a \([k+1,k,2]\) single-parity-check code (SPCC) that provides limited protection against phase-flip errors. Since the SPCC is a special case of the \([n,k,d=2]\) lexicodes introduced by Conway and Sloane~\cite{1057187}, it is natural to seek analogous constructions using higher-distance lexicodes that can encode multiple qubits.

The simplest nontrivial example is the \([5,2,3]\) lexicode, which may be viewed as two overlapping \([3,1,3]\) repetition codes sharing one bit. Generalizing this idea, one can form a family of codes by overlapping \(k\) \([d,1,d]\) repetition codes with \(\ell \le \lfloor d/2 \rfloor\) shared bits, producing a \([k(d-\ell)+\ell,\,k,\,d]\) code. In the limiting case \(k=1\), this reduces to the standard repetition code, while for larger \(k\) the rate improves from \(1/d\) to \(1/(d-\ell)\), allowing up to a twofold enhancement.

After introducing in Sec.~\ref{subsec:overlap} an alternative classical code to replace the repetition code used in the \([[d^{2},1,d]]\) Shor construction, we proposed in Sec.~\ref{sec:construction} three integration strategies. The optimal parameters achieved are
\begin{equation}
\bigl[\!\bigl[\,k\lceil d/2 \rceil^{2}
     + \lfloor d/2 \rfloor\bigl(\lceil d/2 \rceil + 1\bigr)
     + (d-1)\lceil d/2 \rceil,\, k,\, d\,\bigr]\!\bigr].
\end{equation}
All three strategies yield an asymptotic average stabilizer weight of \(4\). A detailed comparison with the surface code and the BB code is provided in Sec.~\ref{subsec:compare}. Some further improved constructions are discussed in Sec.~\ref{sec:further}. Finally, we conclude in Sec.~\ref{sec:conclude}.

\section{Preliminaries}

A quantum stabilizer code that encodes $k$ logical qubits into $n$ physical qubits with code distance $d$ and code rate $k/n$ is denoted by $[[n,k,d]]$. This notation parallels that of a classical linear code, where a code encoding $k$ data bits into $n$ total bits with Hamming distance $d$ is denoted by $[n,k,d]$. In both cases, the code distance corresponds to the minimum weight of an undetectable error.

To fully appreciate the meaning of code distance, it is important to distinguish between two operational settings:
(1) the \textit{forward error correction} (FEC) scenario, in which recovery must be performed solely by the receiver without feedback; and
(2) the \textit{automatic repeat request} (ARQ) scenario, where the receiver can request retransmission upon detecting an error.
In the ARQ setting, any detectable error can be effectively corrected through retransmission, whereas in the FEC setting, one typically adopts the conventional statement that ``an $[[n,k,d]]$ code can correct up to $\lfloor (d-1)/2 \rfloor$ errors and detect up to $d-1$ errors.''

To our knowledge, many students and even some instructors interpret this conventional statement in an overly classical sense: that an error of weight $(d-1)/2 < w \le d-1$ can be detected but not corrected because it is closer to another codeword. However, a more precise formulation is that an $[[n,k,d]]$ code can correct \textit{all} stochastic Pauli errors of weight $w \in {1, \dots, \lfloor (d-1)/2 \rfloor}$ and detect \textit{all} such errors of weight $w \in {1, \dots, d-1}$. Certain higher-weight errors, though not all, may also remain correctable or at least detectable, depending on the specific structure of the code.

To elucidate the structure of both the original Shor code and the proposed enhanced constructions, we first revisit the detailed formulation of the Shor code to prevent potential confusion. The identity operator $I$ and the Pauli operators $X,Y,Z$ are defined as $I = \begin{bmatrix} 1 & 0 \\ 0 & 1  \end{bmatrix}$, $X = \begin{bmatrix} 0 & 1 \\ 1 & 0  \end{bmatrix}$, $Y = \begin{bmatrix} 0 & -\mathrm{i} \\ \mathrm{i} & 0  \end{bmatrix}$, $Z = \begin{bmatrix} 1 & 0 \\ 0 & -1  \end{bmatrix},$ with $\mathrm{i}=\sqrt{-1}$. A subscript denotes the qubit on which the Pauli operator acts; for example, \( X_j \) applies the Pauli \( X \) operator to the \( j \)-th qubit, with \( j \) a non-negative integer.

\subsection{Shor code}

The $[[9,1,3]]$ Shor code is a concatenated code composed of inner and outer repetition codes. It employs two weight-6 stabilizers, \(\{g_0,g_1\}\), for degenerate $Z$ errors and six weight-2 stabilizers, \(\{g_2,\dots,g_7\}\),  for $X$ errors.
Herein, the average stabilizer weight is \(3\).

These stabilizers can be classified into one outer group and three inner groups:
\[
\{(g_0, g_1)\}, \quad \{(g_2, g_3)\}, \quad \{(g_4, g_5)\}, \quad \{(g_6, g_7)\},
\]
where each group consists of two stabilizers forming a repetition code. The first group arises from the outer code, which corrects single phase-flip errors, while the remaining three groups correspond to the inner repetition codes. Single-qubit $Z$ errors can be organized into three degenerate sets, $(Z_{i+0}, Z_{i+1}, Z_{i+2})$ for $i \in \{0, 3, 6\}$, so only two weight-6 stabilizers are required to identify such errors.

The complete syndrome table for all single-qubit errors is presented in Table~\ref{tab:syndrome_913_single}, and selected syndromes for weight-two errors are listed in Table~\ref{tab:syndrome_913_twice}. We observe that once the code can correct single $X$ and $Z$ errors, it can also correct certain weight-two $XZ$ errors, provided the error locations do not overlap. If they do overlap, the combined effect is equivalent to a single $Y$ error.

\begin{table}[ht]
\caption{Syndrome table for single-qubit errors in the $[[9,1,3]]$ Shor code. Degeneracy among $Z$ errors is reflected by identical syndromes.}
\label{tab:syndrome_913_single}
\begin{tabularx}{\linewidth}{ c  c  c  c  c  c  c  c  c }
    \hline\hline
    Error & Syndrome & \hspace{0.4cm} & Error & Syndrome & \hspace{0.4cm} & Error & Syndrome\\
    \hline
    $X_{0}$ & 00100000 & & $Y_{0}$ & 11100000 & & $Z_{0}$ & 11000000\\
    $X_{1}$ & 00110000 & & $Y_{1}$ & 11110000 & & $Z_{1}$ & 11000000\\
    $X_{2}$ & 00010000 & & $Y_{2}$ & 11010000 & & $Z_{2}$ & 11000000\\
    $X_{3}$ & 00001000 & & $Y_{3}$ & 10001000 & & $Z_{3}$ & 10000000\\
    $X_{4}$ & 00001100 & & $Y_{4}$ & 10001100 & & $Z_{4}$ & 10000000\\
    $X_{5}$ & 00000100 & & $Y_{5}$ & 10000100 & & $Z_{5}$ & 10000000\\
    $X_{6}$ & 00000010 & & $Y_{6}$ & 01000010 & & $Z_{6}$ & 01000000\\
    $X_{7}$ & 00000011 & & $Y_{7}$ & 01000011 & & $Z_{7}$ & 01000000\\
    $X_{8}$ & 00000001 & & $Y_{8}$ & 01000001 & & $Z_{8}$ & 01000000\\
    \hline\hline
\end{tabularx}
\end{table}

\begin{table}[ht]
\caption{Syndrome extraction table of weight-two error for the $[[9,1,3]]$ code. Here, we just list a few as it is sufficient to show some errors of weight higher than one is correctable.}
\label{tab:syndrome_913_twice}
\begin{tabularx}{\linewidth}{ c  c  c  c  c }
    \hline
    \hline
    Error & Syndrome & \hspace{3.5cm} & Error & Syndrome\\
    \hline
    $X_{0}Z_{3}$  & 10100000 & $\cdots$ & $Y_{0}Z_{3}$  & 01100000\\
    $X_{1}Z_{3}$  & 10110000 & $\cdots$ & $Y_{1}Z_{3}$  & 01110000\\
    $X_{2}Z_{3}$  & 10010000 & $\cdots$ & $Y_{2}Z_{3}$  & 01010000\\
    $X_{6}Z_{3}$  & 10000010 & $\cdots$ & $Y_{6}Z_{3}$  & 11000010\\
    $X_{7}Z_{3}$  & 10000011 & $\cdots$ & $Y_{7}Z_{3}$  & 11000011\\
    $X_{8}Z_{3}$  & 10000001 & $\cdots$ & $Y_{8}Z_{3}$  & 11000001\\
    \hline
    \hline
\end{tabularx}
\end{table}

The reason we emphasize that certain higher-weight errors can still be corrected or detected is that this distinction significantly influences the estimation of the error threshold at which a particular QECC becomes advantageous. While most QECC analyses treat high-weight errors as ``negligible'' perturbative contributions, largely to avoid the substantial complexity of constructing exhaustive lookup tables for recovery operations, this simplification can lead to an underestimation of a code's actual capability. In many cases, recovery can be governed by simple, rule-based operations; for example, when a specific syndrome is flagged, a corresponding correction can be directly applied. Although a full lookup table scales exponentially as $2^r$ with the number of stabilizers $r$, implementing $r$ simple, syndrome-dependent recovery rules remains tractable.

In the Shor code, the eight stabilizers can be organized into four independent two-bit groups, reducing the decoding task to \((1+3)2^{2}=16\) elementary recovery rules for the full \(2^{8}=256\) syndrome patterns rather than correcting only \(9\times 3\) weight-\(1\) errors. By contrast, treating the code as an unstructured \([[9,1,3]]\) stabilizer code and applying an iterative BP decoder incurs a decoding cost of order \(O(N W \log\!\log N)\)~\cite{Kuo2022}, with \(N=9\) physical qubits and average stabilizer weight \(W=3\).

More complex codes sacrifice this simplicity to achieve higher code rates, which can increase the decoding time, the performance loss and the risk of additional errors during recovery. Since a primary motivation for employing protected qubits is to achieve higher computational performance compared with classical algorithms, we certainly wish to avoid situations where the exponentially growing syndrome lookup table becomes a new bottleneck.

\subsection{Overlapped repetition code}\label{subsec:overlap}

In this subsection, we construct the
\begin{equation}
[k(d-\ell) + \ell,\, k,\, d]
\end{equation}
\textit{overlapped repetition code} for $\ell \le \lfloor d/2 \rfloor$. As indicated by its definition, we start with $k$ independent $[d,1,d]$ repetition codes and merge $k\ell$ qubits into $\ell$ shared qubits, thereby reducing the total number of physical qubits while preserving the overall distance.

Each logical bit-flip operation acting on a single logical qubit has weight $d$. Any operation simultaneously flipping two logical qubits has weight $2(d - \ell) \ge d$, and operations acting on more logical qubits have weight at least $d$. Consequently, the resulting construction indeed realizes a $[k(d-\ell) + \ell,\, k,\, d]$ code.

If this classical code is interpreted as a biased quantum code, the total of $(k(d - \ell - 1) + \ell)$ stabilizers can be grouped into $(k + 2)$ distinct sets. The first $k$ groups correspond to the unshared portions of the initial repetition codes, each containing $(d - \ell - 1)$ weight-\(2\) stabilizers. The next group represents the shared portion, consisting of $(\ell - 1)$ weight-\(2\) stabilizers that retain the repetition-code structure. The final group contains a single weight-$(k + 1)$ stabilizer that connects the $k$ unshared parts with the shared region.

This grouped structure implies that only $k 2^{d - \ell - 1} + 2^{\ell - 1} + 2$ pieces of information are required for decoding, rather than the full syndrome table of size $2^{k(d - \ell - 1) + \ell}$. For moderate $d$ and large $k$, the resulting decoding complexity scales approximately as the $k$-th root of the naive exhaustive approach.

\section{Quantum overlapped-repetition}\label{sec:construction}

Having established the desired classical code, we now consider its integration within the quantum code architecture, either as the outer code, as the inner code, or simultaneously in both roles. Since the construction employing it as the outer code is the most straightforward, we first present that case, followed by the inner-code and dual-application constructions.

\subsection{The outer code case}\label{subsec:outer}

When the $[k(d-\ell) + \ell,\, k,\, d]$ \textit{overlapped repetition code} is employed as the outer code, it can be straightforwardly concatenated with a $[d,1,d]$ inner repetition code, yielding a quantum code with parameters
\begin{equation}
[[d(k(d-\ell) + \ell),\, k,\, d]].
\end{equation}
In the limiting case $k = 1$, this construction reduces to the standard $[[d^2, 1, d]]$ Shor code. Conversely, as $k \rightarrow \infty$, the asymptotic code rate approaches $\tfrac{1}{d(d-\ell)}$. For even $d$ and maximal overlap $\ell = d/2$, the rate becomes $2/d^2$, exactly twice that of the original Shor code. For odd $d$, the corresponding rate is $\tfrac{2}{d(d+1)}$.

The outer layer contributes \(k(d-\ell-1)+\ell\) stabilizers composed entirely of \(X\) operators, one of weight \((k+1)d\) and the remainder of weight \(2d\). The inner layer contributes \((k(d-\ell)+\ell)(d-1)\) stabilizers composed entirely of \(Z\) operators, each of weight \(2\).
The average stabilizer weight is
\begin{equation}
    W_{\mathrm{outer}}
    = \frac{(2(d-\ell)(2d-1)-d)k + 2\ell(2d-1) - d}{((d-\ell)d - 1)k + \ell d},
\end{equation}
which approaches \(W_{\mathrm{outer}} \to 4\) in the limit \(k \gg d \gg 1\).

\subsubsection{Protection}

To verify that the concatenated construction indeed attains distance \(d\), we analyze all error patterns of weight at most \(d\).

Suppose first that \(d' \le d\) \(X\)-type errors occur within a single inner repetition block of distance \(d\). Such an event is uniquely identifiable by the inner code, and thus any distribution of up to \(d\) \(X\) errors across the blocks is detectable.

Conversely, consider \(Z\)-type errors. If \(d' \le d\) \(Z\) errors occur within a single inner repetition block, the inner-code stabilizers ensure that the error has no effect when \(d'\) is even, and induces a single phase flip when \(d'\) is odd. In the worst case, therefore, up to \(d\) such phase flips may be produced across the blocks, each corresponding to an effective \(Z\)-error on one outer-code qubit. Since the outer repetition code has distance \(d\), all such patterns remain detectable.

Taken together, these observations confirm that the concatenated construction has distance \(d\).

\subsubsection{Relation to other codes}

Although Sec.~\ref{subsec:overlap} focuses on the regime \(\ell \le \lfloor d/2 \rfloor\), the underlying idea of overlap has antecedents in other high-rate constructions. One example is the amplitude-damping \(w\) code with parameters \([[(w+1)(w+k),\,k]]\)~\cite{PhysRevA.111.052602,4675715}, which corresponds to choosing \(d = w+1\) and \(\ell = w\). A related perspective appears in bosonic encodings such as the extended binomial code~\cite{hwfz-c6vy}, where excitations within the inner repetition structure can be merged and treated as indistinguishable particles. The essential distinction is that stronger overlap becomes permissible when the physical noise is biased, enabling higher-rate designs that would otherwise violate distance constraints in the unbiased setting.

\subsection{The inner code case}\label{subsec:inner}


In this approach, the inner layer is chosen to be the
\begin{equation}
[\,((k+1)d - 1)(d-\ell) + \ell,\; (k+1)d - 1,\; d\,]
\end{equation}
overlapped repetition code. The outer layer consists of \(k+1\) independent \([d,1,d]\) repetition codes, yielding a concatenated construction with parameters
\begin{equation}
[[\,((k+1)d - 1)(d-\ell) + \ell,\; k,\; d\,]].
\end{equation}
These parameters parallel those of the outer-code variant but may increase the total qubit count by \((d-1)(d-2\ell)\) for fixed logical dimension while preserving the distance. The asymptotic code rate remains twice that of the original Shor construction.

In this setting, the inner layer contributes \((kd + (d-1))(d-\ell-1) + \ell\) stabilizers of \(Z\)-type: one of weight \((k+1)d\) and the remainder of weight \(2\). The outer layer contributes \((k+1)(d-1)\) \(X\)-type stabilizers, of which \(k(d-1)+(d-2)\) have weight \(2(d-\ell)\), and one has weight \(d\).
The average stabilizer weight is
\begin{equation}
    W_{\mathrm{inner}}
    =
    \frac{(2(d-\ell)(2d-1)-d)\,k + 2(d-\ell)(2d-5) + 2d}
         {(d^2 - 1 - d\ell)\,k + (d-1)(d-\ell)\ell},
\end{equation}
which approaches \(W_{\mathrm{inner}}\to 4\) in the limit \(k \gg d \gg 1\).

\subsubsection*{Protection}

Again, to verify that the concatenated construction indeed attains distance \(d\), we analyze all error patterns of weight at most \(d\).

Suppose first that \(d' \le d\) \(X\)-type errors occur within the inner code of distance \(d\). Such an event is uniquely identifiable by the inner code, and hence all configurations with at most \(d\) \(X\) errors are detectable.

For \(Z\)-type errors, the analysis changes because the inner layer is now a single overlapped-repetition code rather than multiple independent repetition codes. This code consists of one shared repetition block of length \(\ell\) and several unshared blocks of length \(d - \ell\). If \(d'\) \(Z\) errors occur within any single block, the effect is trivial when \(d'\) is even and corresponds to a single phase flip when \(d'\) is odd. Thus, in the worst case, up to \(d\) such phase flips may arise across the blocks. If the inner code has parameters \([[k_i(d-\ell)+\ell,\, k_i,\, d]]\), then it contains \(k_i + 1\) blocks, and the outer layer must be designed to account for this additional block.

Consider the choice \(k_i = (k+1)d - 1\). In this setting, \(k\) identical outer repetition codes of distance \(d\) each protect \(d\) blocks, and an additional outer repetition code of distance \(d\) protects the remaining \(d-1\) unshared blocks of length \(d - \ell\) together with the single shared block of length \(\ell\). Since every outer code has distance \(d\), all such error patterns remain detectable.

Taken together, these observations confirm that the concatenated construction indeed has distance \(d\).

\subsection{Both outer and inner codes}\label{subsec:both}

Finally, the overlapping strategy may be applied simultaneously to both the inner and outer layers to further enhance the code rate. As shown in Sec.~\ref{subsec:inner}, the construction requires one additional repetition code beyond the \(k\) identical outer repetition codes. Moreover, Sec.~\ref{subsec:outer} established that these \(k\) outer repetition codes can themselves be merged into a single overlapped-repetition code. Consequently, the outer layer consists of a \([\,k(d-\ell)+\ell,\, k,\, d\,]\) code together with an additional \([\,d,1,d\,]\) repetition code.

These two outer codes require the inner layer to supply \(k_i = k(d-\ell) + \ell + (d-1)\) unshared blocks, along with one shared block. Thus, the inner layer is a \([\,k_i(d-\ell)+\ell,\, k_i,\, d\,]\) overlapped-repetition code. The resulting concatenated quantum code therefore has parameters
\begin{equation}
[[\,k(d-\ell)^2 + \ell(d-\ell) + (d-1)(d-\ell) + \ell,\, k,\, d\,]].
\end{equation}
As $k \rightarrow \infty$, the asymptotic code rate approaches $1/(d-\ell)^2$. Choosing the maximal overlap $\ell = \lfloor d/2\rfloor$ yields an asymptotic rate that is about four times higher than that of the original Shor code.

Among the inner-layer stabilizers, a single stabilizer has weight \(k_i + 1\), while all remaining stabilizers have weight \(2\). For the overlapped-repetition portion of the outer layer, one stabilizer has weight
\begin{equation}
w_k =
\begin{cases}
(k+1)(d-\ell), & \text{if } k \text{ is odd},\\[4pt]
(k+1)(d-\ell) + \ell, & \text{if } k \text{ is even},
\end{cases}
\end{equation}
and the remaining \(k(d-\ell-1) + \ell - 1\) stabilizers each have weight \(2(d-\ell)\). For the additional repetition component of the outer layer, one stabilizer has weight \(d\), and the remaining \(d - 2\) stabilizers each have weight \(2(d-\ell)\). The average stabilizer weight is
\begin{equation}
    W_{\mathrm{both}}
    \approx
    \frac{
        (k_i+k(d-\ell))(2d-2\ell-1)+\cdots
    }{
        (k_i+k)(d-\ell-1)+\cdots
    }.
\end{equation}
In the limit \(k \gg d \gg 1\), this weight converges to \(W_{\mathrm{both}} \to 4\).

\subsubsection*{Protection}

As in the previous subsections, we verify that the concatenated construction attains distance \(d\) by analyzing all error patterns of weight at most \(d\). First, suppose that \(d' \le d\) \(X\)-type errors occur within the inner code of distance \(d\). Such events are uniquely identifiable by the inner layer, ensuring that all configurations with at most \(d\) \(X\) errors are detectable.

For \(Z\)-type errors, the analysis follows the same structure as before. If \(d'\) \(Z\) errors occur within any single repetition block, the effect is trivial when \(d'\) is even and corresponds to a single phase flip when \(d'\) is odd. In the worst case, therefore, up to \(d\) such phase flips may be introduced across the blocks. Take \( k_i = k(d-\ell) + \ell + (d - 1)\). If the inner code has parameters \([[k_i(d-\ell)+\ell,\, k_i,\, d]]\), it contains \(k_i + 1\) blocks, and the outer layer must protect against phase flips on all of them.

In this configuration, the outer overlapped-repetition code of distance \(d\) protects the \(k(d-\ell) + \ell\) blocks of length \(d-\ell\), while an additional outer repetition code of distance \(d\) protects the remaining \(d-1\) unshared blocks of length \(d-\ell\) together with the shared block of length \(\ell\). Since both outer codes have distance \(d\), all such phase-flip patterns remain detectable.

Together, these considerations establish that the concatenated construction achieves distance \(d\).



\subsection{Comparison}\label{subsec:compare}

Three integration strategies are summarized in Table~\ref{tab:compare}.

\begin{table}[ht]
\caption{Comparison of the resulting code parameters. For each construction, we set \(\ell = \lfloor d/2 \rfloor\), which optimizes the code rate. In all cases, the average stabilizer weight approaches \(4\) in the asymptotic regime \(k \gg d \gg 1\).}
\label{tab:compare}
\begin{tabularx}{\linewidth}{ c  c  c }
    \hline\hline
    Code & \hspace{0.4cm} & \([[n,k,d]]\) parameters \\
    \hline
    outer &  & \([[d(k\lceil d/2 \rceil + \lfloor d/2 \rfloor),\, k,\, d]]\) \\[1ex]
    inner &  & \([[((k+1)d-1)\lceil d/2 \rceil + \lfloor d/2 \rfloor,\, k,\, d]]\) \\[1ex]
    both  &  & \([[k\lceil d/2 \rceil^{2} + \lfloor d/2 \rfloor(\lceil d/2 \rceil + 1)+(d-1)\lceil d/2 \rceil,\, k,\, d]]\) \\[1ex]
    \hline\hline
\end{tabularx}
\end{table}
For a direct comparison, taking \(k=d=12\) in the construction that modifies both the inner and outer layers yields parameters \([[540,12,12]]\). This configuration requires more physical qubits than the \([[144,12,12]]\) BB code, but fewer than the \(1452\) (\(2028\)) qubits needed for \(12\) copies of the \([[121,1,11]]\) (\([[169,1,13]]\)) surface code. These values show that the overlapped-repetition Shor construction improves the code rate by a factor in the interval \([2.69,3.76]\) while maintaining a mean stabilizer weight of \(4\). Although this improvement is less dramatic than the \([10.08,14.08]\)-fold increase achieved by the BB code, at the cost of increasing the mean stabilizer weight to \(6\), the overlapped-repetition Shor family preserves a transparent syndrome-to-recovery mapping, thereby avoiding the performance-loss issues associated with iterative decoding.

\section{Further code constructions}\label{sec:further}

\subsection{LDPC design}

The structure identified in Sec.~\ref{subsec:both} suggests that the overlapped-repetition component of the outer layer may be replaced by any efficient classical \([n,k,d]\) code, at the cost of a potentially larger average stabilizer weight. If the chosen outer code is an LDPC code with maximum parity-check weight \(w\), then the corresponding quantum stabilizers have weight at most \(w(d-\ell)\) when \(w\) is even, and at most \(w(d-\ell)+\ell\) when \(w\) is odd.

In this setting, the inner layer must provide \(k_i = n + d - 1\) unshared blocks together with one shared block. The resulting quantum code therefore has parameters
\begin{equation}
[[\,k_i(d-\ell) + \ell,\, k,\, d\,]].
\end{equation}

Suppose the classical \([n,k,d]\) code is asymptotically good, meaning its rate approaches unity in the large-block-length limit. Under this assumption, the resulting quantum construction attains an asymptotic rate approaching \(1/(d-\ell)\), a substantial improvement over the typical \(1/d^2\) scaling.

\subsection{Collective coherent error adapted design}

In addition to stochastic Pauli noise, physical qubits are also subject to collective coherent (CC) errors~\cite{Bravyi2018, Debroy2018, HLRC22, Mrton2023, PhysRevA.90.062317, PhysRevA.93.042340, PhysRevA.111.052602, hwfz-c6vy, qzd2-b2mx}, arising from intrinsic single-qubit Hamiltonians. These errors can periodically reach high weight, and in most settings the precise time of occurrence is unknown, since the receiver may access the encoded state at an arbitrary moment.

A simple and effective method for immune CC errors is to employ constant-excitation encodings. This may be implemented by replacing each inner repetition block with a complement dual-rail map,
\[
(\overline{\ket{0}},\overline{\ket{1}})\ \mapsto\ (\ket{01},\ket{10}),
\]
so that every repetition block of even length maintains a fixed total excitation number. Concretely, the standard repetition basis
\[
(\ket{00\cdots 0},\ \ket{11\cdots 1})
\]
is replaced by
\[
\ket{01\cdots 01},\qquad \ket{10\cdots 10}.
\]

For the construction in Sec.~\ref{subsec:outer}, this substitution is valid whenever \(d\) is even. For the constructions in Sec.~\ref{subsec:inner} and Sec.~\ref{subsec:both}, both \(d-\ell\) and \(\ell\) must be even to ensure that every block has even length. This modification introduces no additional overhead and enhances robustness against CC noise, making it well suited for quantum-memory applications.

\subsection{Bosonic codes}

Interpreting all excitations within a (qubit) block of a repetition code as indistinguishable bosonic excitations of a single oscillator reproduces, in the simplest instances, the (extended) binomial codes~\cite{PhysRevX.6.031006,hwfz-c6vy}.
The same perspective applies directly to the overlapped-repetition constructions developed in this work, yielding a broad family of new bosonic codes obtained by bosonizing the corresponding qubit-based designs.

This approach expands the currently limited set of known bosonic examples, such as cat codes~\cite{PhysRevA.59.2631}, GKP codes~\cite{PhysRevA.64.012310}, and binomial codes, by providing a systematic route to generate additional constructions with tunable rate and distance.

\section{Conclusion}\label{sec:conclude}

We have shown that a simple modification of the standard Shor construction increases its code rate by a factor of four while preserving an asymptotic average stabilizer weight of \(4\). In addition, replacing the overlapped-repetition outer layer with a suitable LDPC code yields a family of quantum codes with asymptotic rate \(2/d\). We have also described a constant-excitation variant that provides protection against collective coherent errors at no additional overhead, as well as a bosonic generalization that extends the construction to oscillator-based codes.

The proposed family (with average stabilizer weight \(4\)) achieves a code rate intermediate between that of the rotated surface code (also with average stabilizer weight \(4\)) and the BB code (average stabilizer weight \(6\)), as detailed in Sec.~\ref{subsec:compare}. Moreover, decoding remains free of the performance degradation typically associated with iterative decoders. A systematic exploration of the trade-off between higher code rate and lower average stabilizer weight is left for future work.


\end{document}